\PassOptionsToPackage{unicode}{hyperref}
\PassOptionsToPackage{hyphens}{url}
\documentclass[
]{article}
\usepackage{lmodern}
\usepackage{amsmath}
\usepackage{ifxetex,ifluatex}
\ifnum 0\ifxetex 1\fi\ifluatex 1\fi=0 
  \usepackage[T1]{fontenc}
  \usepackage[utf8]{inputenc}
  \usepackage{textcomp} 
  \usepackage{amssymb}
\else 
  \usepackage{unicode-math}
  \defaultfontfeatures{Scale=MatchLowercase}
  \defaultfontfeatures[\rmfamily]{Ligatures=TeX,Scale=1}
\fi
\IfFileExists{upquote.sty}{\usepackage{upquote}}{}
\IfFileExists{microtype.sty}{
  \usepackage[]{microtype}
  \UseMicrotypeSet[protrusion]{basicmath} 
}{}
\makeatletter
\@ifundefined{KOMAClassName}{
  \IfFileExists{parskip.sty}{%
    \usepackage{parskip}
  }{
    \setlength{\parindent}{0pt}
    \setlength{\parskip}{6pt plus 2pt minus 1pt}}
}{
  \KOMAoptions{parskip=half}}
\makeatother
\usepackage{xcolor}
\IfFileExists{xurl.sty}{\usepackage{xurl}}{} 
\IfFileExists{bookmark.sty}{\usepackage{bookmark}}{\usepackage{hyperref}}
\hypersetup{
  pdftitle={Kill The Math and Let the Introductory Course Be Born},
  pdfauthor={David Kane},
  hidelinks,
  pdfcreator={LaTeX via pandoc}}
\usepackage[margin=1in]{geometry}
\usepackage{graphicx}
\makeatletter
\def\maxwidth{\ifdim\Gin@nat@width>\linewidth\linewidth\else\Gin@nat@width\fi}
\def\maxheight{\ifdim\Gin@nat@height>\textheight\textheight\else\Gin@nat@height\fi}
\makeatother
\setkeys{Gin}{width=\maxwidth,height=\maxheight,keepaspectratio}
\makeatletter
\def\fps@figure{htbp}
\makeatother
\usepackage{setspace}
\ifluatex
  \usepackage{selnolig}  
\fi
\newlength{\cslhangindent}
\newlength{\csllabelwidth}
\newenvironment{CSLReferences}[2] 
 {
  \setlength{\parindent}{0pt}
  \ifodd #1 \everypar{\setlength{\hangindent}{\cslhangindent}}\ignorespaces\fi
  \ifnum #2 > 0
  \setlength{\parskip}{#2\baselineskip}
  \fi
 }%
 {}
\usepackage{calc}

\title{Kill The Math and Let the Introductory Course Be Born}
\author{David Kane}
\date{}

\begin{document}
\maketitle

\textbf{ABSTRACT}: Our introductory classes in statistics and data
science use too much mathematics. The key causal effect which our
students want our classes to have is to improve their future performance
and opportunities. The more professional their computing skills (in the
context of data analysis), the greater their likely success.
Introductory courses should feature almost no mathematical/statistical
formulas beyond simple algebra.


\hypertarget{introduction}{%
\section{Introduction}\label{introduction}}

In \emph{A Dance with Dragons}, the fifth book in the \emph{The Game of
Thrones} series, Maester Aemon Targaryen advises Jon Snow, newly elected
Lord Commander of the Night's Watch, to ``Kill the boy and let the man
be born.'' Maester Aemon's point was not that there was anything wrong
with a boy's friendships and interests. He was not anti-boy. Instead,
Aemon recognized that those attachments prevented Jon Snow from moving
into manhood, from devoting his full energy to what mattered the most
for his new role.

This article provides similar advice: \emph{Kill the math and let the
introductory course be born.} The mathematical and statistical formulas
found in a typical introductory course are not evil in and of
themselves. Math is fun and beautiful. Yet it is precisely an
over-emphasis on math which has allowed data science courses to thrive,
to meet the informed student demand which statistics departments have
too long ignored. Math is a boyish enthusiasm, unobjectionable when
considered in isolation but inimical to the goals of our students
because it crowds out topics which matter more.

Nolan and Temple Lang (2010) were among the most prominent to recognize
this conflict, insisting that ``Computational literacy and programming
are as fundamental to statistical practice and research as
mathematics.'' In this article, I will discuss the causal effects which
we and our students want our courses to achieve, connect those effects
to some of the specific advice offered by Nolan and Temple Lang (2010),
and recommend replacing large portions of the mathematics we currently
inflict on our students in introductory courses with lessons in
statistical programming. Kaplan (2020) bemoans ``the crawling pace of
the integration of modern computing into the statistics education
curriculum.'' Mathematics ``grips with a dead hand on the undergraduate
{[}statistics{]} curriculum,'' according to Cobb (2015). Let us break
loose.

\hypertarget{goals}{%
\section{Goals}\label{goals}}

Which causal effects do we seek to achieve in our classes? Rumsey
(2002), echoing Cobb (1992) and others, argues for ``statistical
literacy.'' Hardin et al. (2015) want to prepare students to ``think
with data.'' (See also Horton and Hardin (2015).) Aliaga et al. (2005)
used similar terminology, urging us to ``emphasize statistical literacy
and develop statistical thinking'' among our students. No reader will
disagree with these worthy goals. Or, perhaps a better word than
\emph{goals} would be \emph{platitudes}. We need to be much more precise
if we want to make informed choices about how to balance math and
programming in our courses.

De Veaux et al. (2017) discuss guidelines for undergraduate data science
programs. None are objectionable. Yet, in aggregate, they fail to
provide clear metrics for making specific trade-offs. This issue applies
across our curriculum but, for purposes of this article, I will focus on
introductory courses. The (vast?) majority of our students only take one
or two classes in statistics. Consider a collection of causal effects
which an introductory course might have. Rubin (1974) defines a causal
effect as the difference between two potential outcomes. For simplicity,
compare the effect of spending 10 extra hours of student time on
programming versus spending those same 10 hours on mathematics.

We want our students to do well:

\begin{itemize}
\item
  \emph{In the final project for this class.} Such a causal effect is
  only defined, of course, for classes with a final project. For those
  without, however, we can consider the causal effect on final project
  performance if the class were, counterfactually, to have had a final
  project.
\item
  \emph{In research projects for other classes.} As Bray et al. (2014)
  and Mair (2016) explain, reproducible statistical research is
  impossible without basic computer skills. If we want our students to
  do better work in their other classes, we must give them the tools to
  do so.
\item
  \emph{As a summer research intern.} We want out students to perform
  well when they work on summer research for us, and for other faculty.
  As every supervisor knows, independent work outside the classroom is
  very different from the generally well-structured tasks involved in
  class assignments.
\item
  \emph{When working as a summer intern for non-faculty administrators.}
  The first three causal effects are not controversial. Preparing our
  students for academic pursuits --- class projects, summer research ---
  is at the heart of our pedagogical responsibilities. But why should we
  be concerned with how well students do in working for admissions or in
  the investment office? Still, we like our non-faculty colleagues and
  certainly hope our students do good work for them.
\item
  \emph{When working as a summer intern or full time employee outside
  the university.} ``Hold on!'' shout my fellow faculty. ``When did our
  introductory statistics class morph into a seminar at the Office of
  Career Services?'' The answer: When we decided to take our students'
  hopes and dreams seriously. Students know they won't be in school
  forever. Almost all of them will spend the rest of their lives working
  for, and being evaluated by, non-professors at non-academic
  organizations. Problem sets and exams will be left behind, along with
  other childish things. Our students want to do well in the outside
  world. Do we want that? None of us are so misanthropic that we want
  them to do poorly, of course. Yet any honest appraisal of our syllabi
  would suggest that helping them to do well outside of academia is not
  the highest of our priorities.
\item
  \emph{In getting hired in any of the above roles.} However well we
  might prepare students for summer research positions and full time
  employment, that preparation counts for naught if students can't get
  those jobs. How many of us give any thought to increasing our
  students' chances of being hired? Again, we aren't against student
  success \emph{per se}, but very few of us explicitly consider the
  causal effects which our courses have on this outcome. On the margin,
  10 hours spent on coding increases the odds of a job offer much more
  than 10 hours spent on mathematics, however beautiful the mathematics
  might be.
\end{itemize}

Each professor will weigh the relative importance of these causal
effects differently. Some (many?) will put very little weight on the
last two. Your students, however, will put most (almost all?) their
weight on getting a desirable internship/job/career and doing well at
it.

These are not the only outcomes professors (and students) care about.
There is nothing evil about understanding the math behind the sampling
distribution of the mean, the confidence interval for a proportion, a
null hypothesis test, the Central Limit Theorem or the Chi-square
statistic. The causal effect of spending more time on math is to
increase student comprehension of these (important!) theories.

The central reason for the rise of data science courses is that students
recognize, correctly, that taking a typical Data Science 101 course
increases their odds of getting the future they want relative to what
those odds would have been if they had taken Statistics 101. But even
those data science classes don't do as much as they might to improve the
odds.

\hypertarget{an-introductory-course}{%
\section{An Introductory Course}\label{an-introductory-course}}

Even after the improvements in statistics education over the last 20
years, there is still too much math in the typical introductory course.
Consider some of the statistical formulas in De Veaux et al. (2018), a
superb introductory textbook:

\begin{itemize}
\item
  \(\sigma(\hat{p}) = \sqrt{\frac{pq}{n}}\), the sampling distribution
  model for a proportion.
\item
  \(SE(\hat{p}) = \sqrt{\frac{\hat{p}\hat{q}}{n}}\), the standard error
  of an estimated proportion.
\item
  \(\bar{y} \pm t_{n-1}^{*} \times \frac{\sqrt{\frac{\sum(y - \mu)^2}{n}}}{\sqrt{n}}\),
  the one-sample \(t\)-interval for the mean.
\item
  \(t_{n-1} = \frac{\bar{y} - \mu_0}{\frac{\sqrt{\frac{\sum(y - \mu)^2}{n}}}{\sqrt{n}}}\),
  the one-sample \(t\)-test for the mean.
\item
  \(SD(\hat{p_1} - \hat{p_2}) = \sqrt{\frac{p_1 q_1}{n_1} + \frac{p_2 q_2}{n_2}}\),
  the sampling distribution model for a difference between proportions.
\item
  \(SE(\hat{p_1} - \hat{p_2}) = \sqrt{\frac{\hat{p_1}\hat{q_1}}{n_1} + \frac{\hat{p_2}\hat{q_2}}{n_2}}\),
  a two-proportion \(z\)-interval.
\item
  \(z = \frac{(\hat{p_1} - \hat{p_2}) - 0}{\sqrt{\frac{\hat{p_1}\hat{q_1}}{n_1} + \frac{\hat{p_2}\hat{q_2}}{n_2}}}\),
  a two-proportion \(z\)-test.
\item
  \(t = \frac{(\bar{y_1} - \bar{y_2}) - (\mu_1 - \mu_2)}{\sqrt{\frac{s^2_1}{n_1} + \frac{s^2_2}{n_2}}}\),
  a sampling distribution model for the difference between two means.
\item
  \(t = \frac{(\bar{y_1} - \bar{y_2}) - \Delta_0}{\sqrt{\frac{s^2_1}{n_1} + \frac{s^2_2}{n_2}}}\),
  a two-sample \(t\)-test for the difference between means.
\item
  \(t_{n-1} = \frac{\bar{d} - \Delta_0}{\frac{s_d}{\sqrt{n}}}\), the
  paired \(t\)-test.
\item
  \(\chi^2 = \sum \frac{(Obs - Exp)^2}{Exp}\), the Chi-square statistic.
\item
  \(t = \frac{b_1 - \beta_1}{\frac{s_e}{\sqrt{n - 1} s_x}}\), a sampling
  distribution for regression slopes.
\item
  \(VIF = \frac{1}{(1 - R^2_j)}\), the variance inflation factor.
\item
  \(SE(\hat{\mu}_\nu) = \sqrt{SE^2(b_1) \times (x_\nu - \bar{x})^2 + \frac{s^2_e}{n}}\),
  confidence interval for a predicted mean value.
\item
  \(SE(\hat{y}_\nu) = \sqrt{SE^2(b_1) \times (x_\nu - \bar{x})^2 + \frac{s^2_e}{n} + s^2_e}\),
  the prediction interval for an individual.
\end{itemize}

I use De Veaux et al. (2018) for this example, not because it is
excessively mathematical, but because it represents the very best of
accepted practice, including a solid introduction to randomization-based
inference. With regard to the changes made for the current edition, the
authors insist (p.~xi) that ``all of these enhancements follow the new
Guidelines for Assessment and Instruction in Statistics Education
(GAISE) 2016 report'' resulting in ``a course that is more aligned with
the skills needed in the 21st century, one that focuses even more on
statistical thinking and makes use of technology in innovative ways,
while retaining core principles and topic coverage.'' And who is better
positioned than second author Paul Velleman, a co-author of both the
GAISE 2005 and GAISE 2016 reports, to make such a claim?

The central problem with curriculum design, as with writing, is knowing
what to cut. De Veaux et al. (2018) explicitly refuse that hard choice,
noting that ``Many first statistics courses serve wide audiences of
students who need these skills for their own work in disciplines where
traditional statistical methods are, well, traditional. So we have not
reduced our emphasis on the concepts and methods you expect to find in
our texts.''

In other words, despite decades of committee meetings and reports, these
authors present --- and expect their students to learn --- about the
same amount of mathematics as they would have last century. I urge a
different approach.

Consider \href{https://www.davidkane.info/files/gov_1005.html}{Gov 50:
Data}, a Harvard course which I have taught for the last 4 semesters.
(Kane (2020a)) This is an introductory statistics course, focused on
meeting the needs of political science majors but open to the entire
university. In the most recent semester, we had 120 students, ranging in
level from first years through masters students, with the typical
student having no background in statistics or programming. We use an
open-source textbook (Kane (2020b)) created specifically for the course.
One motto for the course is that, even though everyone is welcome, the
target audience is ``poets and philosophers,'' students whose main
academic focus is elsewhere but who recognize that some basic data
analysis skills will be helpful, in both their other courses and after
graduation. The course is designed to be the answer to the following
question: ``If I only want to devote one course to learning skills which
are valued outside the university, which course should I take?''

I never discuss the formulas which De Veaux et al. (2018) spend so much
time on. I never ask a question about them in problem sets or exams. I
devote all that time, and more, toward improving students' computational
skills. My advice:

\begin{itemize}
\item
  \emph{Use Git and Github}. See Bryan (2018), Fiksel et al. (2019) and
  Beckman et al. (2020) for guidance. It is possible to write an essay
  using just a typewriter. It is possible to write code without using
  source control. Neither is professional. Although Git/Github are the
  most popular tools today, that will change over time. The important
  point is not the exact tool. The key is that students should, as much
  as possible, work with professional tools in a professional fashion.
  No employer wants to teach your students about source control.
\item
  \emph{Use an open source programming language}. For introductory
  classes, R (R Core Team (2019)) is the best choice. See
  Çetinkaya-Rundel and Bray (2012), Carson and Basiliko (2016), Silva
  and Moura (2020) and Long and Turner (2020) and for discussion. Python
  can also work, especially if some prior programming experience is a
  prerequisite for the class. These are the statistical programming
  languages that employers want our students to know.
\item
  \emph{Teach randomization-based inference}. Cobb (2007) provides some
  background on why we should prefer such an approach. See Wardrop
  (1995) for an early approach and Ernst (2004) for an overview. Ismay
  and Kim (2019) is an excellent textbook, suitable for use in an
  introductory class. Although there is evidence (Hildreth et al.
  (2018)) that a simulation-based curriculum improves student
  understanding, an even bigger advantage, relative to the traditional
  approach, is the excuse it offers for programming practice. The more
  code that students write, the better they will become at coding.
\item
  \emph{Flip the classroom}. Nielson et al. (2018) demonstrate that
  students learn more in a flipped classroom. Perhaps even more
  important, however, is that a flipped classroom allows for more time
  to be spent in supervised practice. Every minute spent confronting
  real data --- importing, shaping, visualizing, modeling --- is a
  minute well-spent. Can the same be said for every minute of every
  lecture? If students will only devote X hours per week to an
  introductory course, the best way to ensure that they got as much
  practice as possible is to devote class time to that practice. You
  learn soccer with the ball at your feet. You learn data science with
  your hands on the keyboard.
\item
  \emph{Cold-call}. There is no better way to ensure that students are
  engaged in a class than to call on them at random. See Dallimore et
  al. (2012) for a literature review and Lemov (2015) for practical
  details.
\item
  \emph{Require that student work be reproducible}. See Bray et al.
  (2014) for motivation and Baumer et al. (2014) for advice. If your own
  work is reproducible (and I assume that it is), why wouldn't you
  demand the same level of rigor from your students?
\item
  \emph{Require that student work be public}. Students are somewhat
  diligent with the work that only you will see. They take work which
  others will see much more seriously. Requiring that student work be
  public is the easiest way, or at least the most pleasant way, to cause
  students to work harder. And the harder they work, the more likely
  they are to learn something. Naive observers might fear that
  government regulations, like the Family Educational Rights and Privacy
  Act (FERPA), prevent faculty from requiring work be made public. This
  is not true. See Ramirez (2009) for details. FERPA, and similar
  regulations, apply to the records --- grades, comments, et cetera ---
  which faculty create. You can't make those public. You may require
  that student work be public. See bit.ly/1005\_projects for several
  hundred projects completed by my students.
\item
  \emph{Require solo final projects}. See Ledolter (1995) for background
  and discussion of final (or research) projects in statistics classes.
  Imagine that you are trying to decide between two otherwise similar
  candidates for a summer research position. The first has created an
  impressive final project, written in R and hosted on Github, for which
  all the analysis is reproducible. The second has an impressive grasp
  of the mathematics behind a \(t\)-test. Who would you hire? Outside of
  academia, where almost no one uses much mathematics, the first student
  has a big advantage, just as Horton (2015) reports. Projects should be
  solo because otherwise students will divide-and-conquer the work,
  often with one doing all the writing and the other all the coding,
  which is not the outcome we want.
\end{itemize}

My introductory course does all these things. The combined causal
effects of these requirements is to significantly increase my students'
odds of achieving the futures they want, relative to what those odds
would have been if my course were more mathematical. Computational
skills are more relevant than mathematical understanding to their future
success.

A similar course would work just as well outside of Harvard. First,
there is an extensive overlap between the bottom quarter of the Harvard
distribution of student ability and the student bodies of other top
tercile schools. Second, weaker students are even more interested in
learning skills which employers value. With less talented students, I
would follow exactly the same approach while, perhaps, cutting back on
the total number of topics covered.

\hypertarget{conclusion}{%
\section{Conclusion}\label{conclusion}}

Horton (2015) writes:

\begin{quote}
Anecdotal reports have indicated that statistics undergraduates are at a
competitive disadvantage relative to undergraduate computer science (CS)
degree holders for entry level positions. Such positions tend to have
data-related skills at the core of the job descriptions. At present,
many CS students tend to be able to perform computations with data more
easily than their statistics equivalents. This should not be allowed to
continue.
\end{quote}

``Allowed by whom?'' one might ask. The central problem is that, even a
decade after Nolan and Temple Lang (2010) made a similar point, we have
done too little. Nolan and Temple Lang (2010) quoted Friedman (2001):

\begin{quote}
Computing has been one of the most glaring omissions in the set of tools
that have so far defined Statistics. Had we incorporated computing
methodology from its inception as a fundamental statistical tool (as
opposed to simply a convenient way to apply our existing tools) many of
the other data related fields would not have needed to exist. They would
have been part of our field.
\end{quote}

Indeed. But it is never too late to change course. Do you need to make
the full leap described above next semester? No.~Start slowly. Remove 10
hours of time spent on math and add 10 hours more on programming.

Friedman (2001) also suggested that ``We may have to moderate our
romance with mathematics.'' There is no ``may'' about it. To make room
for computation, we must ditch something. The most obvious something is
mathematics. We should stop using mathematics beyond algebra in our
introductory courses. No question on a problem set or exam should
require the use of a formula. We need radical surgery if our courses are
to have the causal effects which our students most want. My message to
instructors: Stop using mathematics, despite your fond memories of how
it helped you in the past, and, from this point forward, use only
computers. \emph{Kill the math and let the introductory course be born.}

\emph{Acknowledgment}: I thank Joe Blitzstein, Mike Parzen and Liberty
Vittert for useful discussions. Special thanks to three anonymous
reviewers and to as associate editor of the \emph{Journal of Statistics
Education} for their comments, and to an anonymous reviewer and an
editor at the \emph{Harvard Data Science Review}.

\hypertarget{references}{%
\section*{References}\label{references}}
\addcontentsline{toc}{section}{References}

\hypertarget{refs}{}
\begin{CSLReferences}{1}{0}
\leavevmode\hypertarget{ref-aliaga2005guidelines}{}%
Aliaga, M., Cobb, G., Cuff, C., Garfield, J., Gould, R., Lock, R.,
Moore, T., Rossman, A., Stephenson, B., Utts, J., Velleman, P., and
Witmer, J. (2005), \emph{Guidelines for assessment and instruction in
statistics education ({GAISE}): {C}ollege {R}eport}, \emph{American
Statistical Association}.

\leavevmode\hypertarget{ref-Baumer2014}{}%
Baumer, B., Çetinkaya-Rundel, M., Bray, A., Loi, L., and Horton, N. J.
(2014), {``R markdown: Integrating a reproducible analysis tool into
introductory statistics,''} \emph{Technology Innovations in Statistics
Education}, University of California: Berkeley Electronic Press, 8.

\leavevmode\hypertarget{ref-beckman2020}{}%
Beckman, M. D., Çetinkaya-Rundel, M., Horton, N. J., Rundel, C. W.,
Sullivan, A. J., and Tackett, M. (2020), {``Implementing version control
with git as a learning objective in statistics courses.''}

\leavevmode\hypertarget{ref-Bray2014}{}%
Bray, A., Çetinkaya-Rundel, M., and Stangl, D. (2014), {``Five concrete
reasons your students should be learning to analyze data in the
reproducible paradigm,''} \emph{Chance}, Abingdon: Taylor \& Francis
Ltd., 27, 53.

\leavevmode\hypertarget{ref-Bryan2018}{}%
Bryan, J. (2018), {``Excuse me, do you have a moment to talk about
version control?''} \emph{The American Statistician}, Taylor \& Francis,
72, 20.

\leavevmode\hypertarget{ref-Carson2016}{}%
Carson, M. A., and Basiliko, N. (2016), {``Approaches to r education in
canadian universities,''} \emph{F1000 research}, England: Faculty of
1000 Ltd, 5, 2802.

\leavevmode\hypertarget{ref-cobb1992heeding}{}%
Cobb, G. (1992), {``Heeding the call for change: Suggestions for
curricular action,''} \emph{Teaching Statistics}, 22, 3--43.

\leavevmode\hypertarget{ref-Cobb2007}{}%
Cobb, G. (2007), {``The introductory statistics course: A ptolemaic
curriculum?''} University of California: eScholarship.

\leavevmode\hypertarget{ref-Cobb2015}{}%
Cobb, G. (2015), {``Mere renovation is too little too late: We need to
rethink our undergraduate curriculum from the ground up,''} \emph{The
American Statistician}, Taylor \& Francis, 69, 266--282.
\url{https://doi.org/10.1080/00031305.2015.1093029}.

\leavevmode\hypertarget{ref-CR2012}{}%
Çetinkaya-Rundel, M., and Bray, A. (2012), {``Integrating {R} into
introductory statistics.''} Joint Statistical Meetings.

\leavevmode\hypertarget{ref-Dallimore2012}{}%
Dallimore, E. J., Hertenstein, J. H., and Platt, M. B. (2012), {``Impact
of cold-calling on student voluntary participation,''} \emph{Journal of
management education}, Los Angeles, CA: SAGE Publications, 37, 305--341.

\leavevmode\hypertarget{ref-deveaux2017}{}%
De Veaux, R. D., Agarwal, M., Averett, M., Baumer, B. S., Bray, A.,
Bressoud, T. C., Bryant, L., Cheng, L. Z., Francis, A., Gould, R., Kim,
A. Y., Kretchmar, M., Lu, Q., Moskol, A., Nolan, D., Pelayo, R.,
Raleigh, S., Sethi, R. J., Sondjaja, M., Tiruviluamala, N., Uhlig, P.
X., Washington, T. M., Wesley, C. L., White, D., and Ye, P. (2017),
{``Curriculum guidelines for undergraduate programs in data science,''}
\emph{Annual Review of Statistics and Its Application}, 4, 15--30.
\url{https://doi.org/10.1146/annurev-statistics-060116-053930}.

\leavevmode\hypertarget{ref-deveaux18}{}%
De Veaux, R., Velleman, P., and Bock, D. (2018), \emph{Intro stats},
Boston: Pearson.

\leavevmode\hypertarget{ref-Ernst2004}{}%
Ernst, M. D. (2004), {``Permutation methods: A basis for exact
inference,''} \emph{Statistical Science}, Institute of Mathematical
Statistics, 19, 676--685.

\leavevmode\hypertarget{ref-Fiksel2019}{}%
Fiksel, J., Jager, L. R., Hardin, J. S., and Taub, M. A. (2019),
{``Using {G}ithub {C}lassroom to teach statistics,''} \emph{Journal of
Statistics Education}, Taylor \& Francis, 27, 110--119.

\leavevmode\hypertarget{ref-Friedman2001}{}%
Friedman, J. H. (2001), {``The role of statistics in the data
revolution?''} \emph{International Statistical Review}, Wiley, 69,
5--10.

\leavevmode\hypertarget{ref-hardin2015}{}%
Hardin, J., Hoerl, R., Horton, N. J., Nolan, D., Baumer, B., Hall-Holt,
O., Murrell, P., Peng, R., Roback, P., Temple Lang, D., and Ward, M. D.
(2015), {``Data science in statistics curricula: Preparing students to
{`{T}hink with {D}ata'},''} \emph{The American Statistician}, Taylor \&
Francis, 69, 343--353.
\url{https://doi.org/10.1080/00031305.2015.1077729}.

\leavevmode\hypertarget{ref-Hildreth18}{}%
Hildreth, L. A., Robison-Cox, J., and Schmidt, J. (2018), {``Comparing
student success and understanding in introductory statistics under
consensus and simulation-based curricula,''} \emph{Statistics Education
Research Journal}, 17, 103--120.

\leavevmode\hypertarget{ref-Horton20015co}{}%
Horton, N. J. (2015), {``Challenges and opportunities for statistics and
statistical education: Looking back, looking forward,''} \emph{The
American Statistician}, Taylor \& Francis, 69, 138--145.
\url{https://doi.org/10.1080/00031305.2015.1032435}.

\leavevmode\hypertarget{ref-horton2015}{}%
Horton, N. J., and Hardin, J. S. (2015), {``Teaching the next generation
of statistics students to {`{T}hink with {D}ata'}: Special issue on
statistics and the undergraduate curriculum,''} \emph{The American
Statistician}, Taylor \& Francis, 69, 259--265.
\url{https://doi.org/10.1080/00031305.2015.1094283}.

\leavevmode\hypertarget{ref-ModernDive}{}%
Ismay, C., and Kim, A. Y. (2019), \emph{Statistical inference via data
science: A {M}odern{D}ive into {R} and the {T}idyverse}, Chapman \&
hall/CRC the r series, CRC Press.

\leavevmode\hypertarget{ref-kane2020}{}%
Kane, D. (2020a), {``Syllabus for {G}ov 1005: {D}ata,''} Zenodo.
\url{https://doi.org/10.5281/zenodo.3766268}.

\leavevmode\hypertarget{ref-ppbds}{}%
Kane, D. (2020b), \emph{Preceptor's {P}rimer for {B}ayesian {D}ata
{S}cience}, Zenodo. \url{https://doi.org/10.5281/zenodo.3766374}.

\leavevmode\hypertarget{ref-kaplan2020}{}%
Kaplan, D. (2020), {``StatPREP: Living in interesting times for
introductory statistics education,''} \emph{Amstat News}, American
Statistical Association, 36.

\leavevmode\hypertarget{ref-LedolterJohannes1995PiIS}{}%
Ledolter, J. (1995), {``Projects in introductory statistics courses,''}
\emph{The American Statistician}, Taylor \& Francis Group, 49, 364--367.

\leavevmode\hypertarget{ref-Lemov2015}{}%
Lemov, D. (2015), \emph{Teach like a champion 2.0 : 62 techniques that
put students on the path to college}, San Francisco: Jossey-Bass.

\leavevmode\hypertarget{ref-Long2020}{}%
Long, J. D., and Turner, D. (2020), {``Applied r in the classroom,''}
\emph{Australian Economic Review}, Wiley Subscription Services, Inc, 53,
139--157.

\leavevmode\hypertarget{ref-mair16}{}%
Mair, P. (2016), {``Thou shalt be reproducible! A technology
perspective,''} \emph{Frontiers in Psychology}, 7, 1079.
\url{https://doi.org/10.3389/fpsyg.2016.01079}.

\leavevmode\hypertarget{ref-Nielson18}{}%
Nielson, P. L., Bean, N. W. B., and Larsen, R. A. A. (2018), {``The
impact of a flipped classroom model of learning on a large undergraduate
statistics class,''} \emph{Statistics Education Research Journal}, 17,
121--140.

\leavevmode\hypertarget{ref-nolan2010}{}%
Nolan, D., and Temple Lang, D. (2010), {``Computing in the statistics
curricula,''} \emph{The American Statistician}, Taylor \& Francis, 64,
97--107. \url{https://doi.org/10.1198/tast.2010.09132}.

\leavevmode\hypertarget{ref-Rcitation}{}%
R Core Team (2019), \emph{R: A language and environment for statistical
computing}, Vienna, Austria: R Foundation for Statistical Computing.

\leavevmode\hypertarget{ref-ferpa}{}%
Ramirez, C. A. (2009), \emph{FERPA clear and simple: The college
professional's guide to compliance}, The jossey-bass higher and adult
education series FERPA clear and simple, San Francisco: Jossey-Bass.

\leavevmode\hypertarget{ref-Rubin1974}{}%
Rubin, D. B. (1974), {``Estimating causal effects of treatments in
randomized and nonrandomized studies,''} \emph{Journal of Educational
Psychology}, American Psychological Association, 66, 688--701.

\leavevmode\hypertarget{ref-rumsey2002}{}%
Rumsey, D. J. (2002), {``Statistical literacy as a goal for introductory
statistics courses,''} \emph{Journal of Statistics Education}, Taylor \&
Francis, 10. \url{https://doi.org/10.1080/10691898.2002.11910678}.

\leavevmode\hypertarget{ref-silva2020}{}%
Silva, H. A. da, and Moura, A. S. (2020), {``Teaching introductory
statistical classes in medical schools using RStudio and r statistical
language: Evaluating technology acceptance and change in attitude toward
statistics,''} \emph{Journal of Statistics Education}, Taylor \&
Francis, 0, 1--8. \url{https://doi.org/10.1080/10691898.2020.1773354}.

\leavevmode\hypertarget{ref-wardrop95}{}%
Wardrop, R. L. (1995), \emph{Statistics: Learning in the presence of
variation}, William C. Brown.

\end{CSLReferences}

\end{document}